\documentclass[preprint,superscriptaddress]{revtex4}
\usepackage[dvips]{graphicx}
\usepackage{setspace}
\usepackage{amssymb,amsfonts,amsmath}

\begin{document}
\begin{center}
{\large \bf Dynamics of the bacterial flagellar motor with multiple stators\\}
\vspace{0.5cm}
Giovanni Meacci and Yuhai Tu$^{*}$\\
\vspace{0.3cm} IBM T. J. Watson Research Center\\ P.O. Box 218, Yorktown
Heights, NY 10598\\
$^*$Corresponding author (email: yuhai@us.ibm.com)\\
\end{center}
\vspace{2.5cm}
\begin{center}
{\bf Abstract}
\end{center}
The bacterial flagellar motor drives the rotation of flagellar filaments and enables many species of bacteria to swim. Torque is generated by interaction of stator units, anchored to the peptidoglycan cell wall, with the rotor. Recent experiments [Yuan, J. \& Berg, H.~C. (2008) \emph{PNAS} 105, 1182-1185] show that near zero load the speed of the motor is independent of the number of stators. Here, we introduce a mathematical model of the motor dynamics that explains this behavior based on a general assumption that the stepping rate of a stator depends on the torque exerted by the stator on the rotor. We find that the motor dynamics can be characterized by two time scales: the moving-time interval for the mechanical rotation of the rotor and the waiting-time interval determined by the chemical transitions of the stators. We show that these two time scales depend differently on the load, and that their crossover provides the microscopic explanation for the existence of two regimes in the torque-speed curves observed experimentally. We also analyze the speed fluctuation for a single motor using our model. We show that the motion is smoothed by having more stator units. However, the mechanism for such fluctuation reduction is different depending on the load. We predict that the speed fluctuation is determined by the number of steps per revolution only at low load and is controlled by external noise for high load. Our model can be generalized to study other molecular motor systems with multiple power-generating units.
\newpage

\begin{singlespace}
The swimming motion of bacterium \emph{Escherichia coli} is propelled by the concerted rotational motion of its flagellar filaments \cite{BA73, B03}. Each filament ($\sim 10\mu$m long) is driven by a rotatory motor embedded in the cell wall, with a angular speed of the order of $100$ Hz \cite{B03}. The motor has one rotor and multiple stators in a circular ring-like structure roughly $45$nm in diameter \cite{TFXD06}. The stators are attached to the rigid peptidoglycan cell wall and the spinning of the rotor drives the flagellar filament through a short hook (see \cite{TFXD06} for a 3D reconstruction and Fig. 1(a) for a 2D sketch of the rotor-stator spatial arrangement). The rotor is composed of a ring of $\sim 26$ FliG proteins and each stator has four copies of proteins MotA and two copies of proteins MotB, forming two proton-conducting transmembrane channels. A flow of protons (or, in some alkalophilic and marine \emph{Vibrio} species of bacteria, sodium ions), due to electrochemical gradients across the channels, causes conformational changes of the stator proteins that generate force on the rotor through electrostatic interaction between MotA and protein FliG \cite{KB01}. The work per unit charge that a proton can do in crossing the cytoplasmic membrane through the proton channel is called the ``proton-motive force'' (\emph{pmf}).

At any given time, a stator is engaged with one of the 26 FliG monomers on the FliG ring as the duty ratio of the flagellar motor is close to unity\cite{RBB00}. Presumably, the passage of protons switches the stator to be engaged with the next FliG monomer on the FliG ring along the direction of rotation, stretching the link between the stator and the rotor. The subsequent relaxation process rotates the rotor and the attached load towards the new equilibrium position. This can give rise to a step-like motion, characterized by advances of the rotor followed by waiting periods. The molecular details of the flagellar motor has been the subject of intense research\cite{B03} and the step-like motion was recently demonstrated by direct observation \cite{SRLYHIB05} for a sodium-powered motor at very low \emph{pmf}, but a general understanding of the stepping dynamics of a single flagellar motor is still lacking.

The torque-speed dependence is the key characteristics of the motor\cite{B03,CB00a,SHHI03}. The measured torque-speed curve (see Supporting Information (SI)) for bacterial flagellar motor shows two distinctive regimes. From its maximum value $\tau_{max}$ at stall (zero angular velocity), the torque first falls slowly (by roughly $10\%$) as angular velocity increases at up to a large fraction ($\approx 60 \%$) of the maximum velocity, forming a plateau in the torque-speed curve. Then the torque starts to decrease quickly with increasing angular velocity, eventually approaches zero at the maximum velocity. For \emph{E. coli} at room temperature under physiologically relevant conditions, the maximum angular velocity is $\approx 300$Hz and the estimated maximum torque ranges from $4700$~pN-nm\cite{BB97} to $1400$~pN-nm \cite{FRB03,RLCLAB06}.

A few mathematical models \cite{L88,B93,MCB89,BPGEVPB99,WC00,S03} have been proposed to explain various aspects of the observed torque-speed characteristics based on assumptions about details of the electrostatic interaction between the stators and the rotor. In a more general approach, recent work by Xing et al. \cite{XBBO06} has sought to understand the mechanism for the torque-speed curve characteristics without assuming a detailed description of the energy-transduction process. Their model can reproduce the observed torque-speed curve characteristics, and a set of general conditions to explain the observed torque-speed characteristics were suggested. However, the model by Xing et al. does not exhibit the correct behavior at low load. In their model, the maximum velocities depend inversely on the number of stators (see Supporting Text in \cite{XBBO06}), whereas a recent experiment \cite{YB08} shows that near zero load the velocity of the motor is independent of the number of stators.

Here, we aim at understanding both the torque-speed relationship and the individual motor dynamics by using a simple model describing the rotor's mechanical motion and the stator's stepping probability.
In our model, the stepping rate of a stator depends on the force between the stator and the rotor, in analogy to the Huxley model for Myosin\cite{H57}. Specifically, ``negative" force between a stator and its attached FliG monomer in the direction opposite to the rotation of the motor leads to a larger stepping rate for the stator. Under this general assumption, we find that the maximum velocity at (near) zero load in our model is mostly determined by the maximum stepping rate, independent of the number of stators, in agreement with the recent experiment by Yuan and Berg\cite{YB08}. Microscopically, the motor dynamics follows a repeated moving and waiting pattern characterized by two time scales: the moving-time interval $t_{m}$ associated with the (mechanical) rotation of the rotor and the waiting-time interval $t_{w}$ determined by the (chemical) transition of the stator. We find that $t_m$ and $t_w$ depend differently on the load and their crossover provides a natural explanation for the observed two regimes of the torque-speed curve. The fluctuation of the motor rotation are also studied in our model. We show that the sources of the motor speed fluctuation are totally different in the high and low load regimes and that the number of steps per revolution can only be extracted from the analysis of motor speed fluctuation in the low load limit.

\section{Model}
In Fig. 1(a), a schematic representation of a flagellar motor (rotor and stators) is shown. Each stator has two force-generating subunits symbolized by the light-blue and the red springs. The two force units of a stator interact with the FliG ring (rotor) in a hand-over-hand fashion as illustrated in Fig. 1(b), analogous to the way kinesin proteins interact with microtubules \cite{AFB03, YTVS04}. The switching of hands (force-generating unit) represents the energy-assisted transition when one hand releases its attachment and the other hand establishes its interaction with the FliG ring (rotor). The forces between the FliG ring and the stators drive the rotation of the rotor. In Fig. 1(c), the corresponding sequence of this hand-over-hand motion is shown in the energy landscape. The physical motion (solid arrow) of the rotor (green circle) is governed by its interaction potential with the engaged FliG. The hand-switch transition (dotted arrow) corresponds to a shift of the potential energy in the direction of motor rotation by angle $\delta_0$ and the subsequent motor motion is governed by this new potential until the next switch. Microscopically, the shift angle could be different for the front hand and the back hand (with respect to the direction of the motor rotation); here for simplicity $\delta_0$ is a constant. Commensurate with the periodicity $\delta=2\pi/26$ of the FliG ring, we should have $2\delta_0=m\delta$ with a small integer $m$. In this paper we choose $m=1$ for simplicity.

Due to the small Reynolds number, the dynamics of the rotor angle $\theta$ and the load angle $\theta_L$ are over-damped
and can be described by the following Langevin equations:
\begin{eqnarray}\label{eqrotor}
\xi_{R} \frac{d\theta}{dt}=-\frac{\partial}{\partial\theta}\sum_{i=1}^{N}V(\theta-\theta_i^S)- F(\theta-\theta_{L})
+\sqrt{2k_{B}T \xi_{R}} \alpha(t),\label{e1}
\end{eqnarray}
\begin{eqnarray}\label{eqload}
\xi_{L} \frac{d\theta_{L}}{dt}= F(\theta-\theta_{L})+\sqrt{2k_{B}T \xi_{L}} \beta(t),\label{e2}
\end{eqnarray}
where $\xi_{R}$ and $\xi_{L}$ are the drag coefficients for the rotor and the load respectively, and $N$ is the total number of stators in the motor.
$V$ is the interaction potential between the rotor and the stator. $V$ depends on the relative angular coordinates $\Delta\theta_{i}=\theta-\theta_{i}^{S}$, where $\theta_{i}^{S}$ is the internal coordinate of the stator $i$. $\theta_i^{S}$ increases by $\delta_0$ when the stator switches hands. This discrete change in $\theta_i^S$ is called a jump of the stator in this paper. The load is coupled to the rotor via a nonlinear spring described by a function $F$, which can be determined from the hook spring compliance measurement of ref. \cite{BBB89} (see Fig. S1 in SI). The last terms in Eq. (\ref{e1}-\ref{e2}) are stochastic forces acting on the rotor and on the load, with $k_{B}$ the Boltzmann constant, $T$ the absolute temperature, and $\alpha(t)$ and $\beta(t)$ independent white noise fluctuations of unity intensity.

The dynamics of the stator $i$ is governed by the transition probability for the discrete jump of its internal variable $\theta_i^{S}$ during the time interval $t$ to $t+\Delta t$: $P_i(\theta_i^S\rightarrow\theta_i^S+\delta_0)$. In this paper, $P_i$ is assumed to depend on the torque generated by the $i$'th stator $\tau_i\equiv -V'(\Delta\theta_i)$, which depends on the relative angle $\Delta\theta_i$:
\begin{equation}
P_i(\theta_i^S\rightarrow\theta_i^S+\delta_0)=r(\tau_i)\Delta t=k(\Delta\theta_i)\Delta t.
\label{e3}
\end{equation}
The specific form of the jumping rate $r(\tau_i)$ (or $k(\Delta\theta_i)$) is unknown. We assume it to be a decreasing function of $\tau_i$, with the stator stepping rate being higher when $\tau_i$ is negative ($\tau_i<0, \Delta\theta_i>0$) than when $\tau_i$ is positive ($\tau_i>0, \Delta\theta_i<0$).

Fig. 1(d) illustrates the motor dynamics, where
the rotor (green circle) is either pulled forward or dragged backward by individual stators (purple circles) depending on their relative coordinates with respect to the rotor. The stator coordinate changes by jumping forward by $\delta_0$ with a probability rate that is a function of its relative coordinate. For simplicity, we set the potential function $V$ to be a $V$-shaped function: $V(\Delta\theta)=\tau_0|\Delta\theta|$, and the torque from a single stator is $\tau_0$ with its sign depending on whether the stator is pulling ($\Delta\theta<0$) or dragging ($\Delta\theta>0$). Correspondingly, the stator jumping rate depends on the sign of the force: $k(\Delta\theta<-\delta_c)=0$, $k(-\delta_c<\Delta\theta<0)=k_+$, $k(\Delta\theta>0)=k_-(>k_+)$ as illustrated in Fig. 1(e). A cutoff angle $\delta_c$ is introduced to prevent run-away stators. 
Quantitatively, we use $\tau_0=505$pN-nm, $\xi_R=0.02$pN-nm-s-rad$^{-1}$, $k_{+}=12000$s$^{-1}$, $k_{-} =2 k_{+}$, $\delta_c=\delta_0$ in this paper unless otherwise stated. The load $\xi_L$ varies from $0.002-50$pN-nm-s-rad$^{-1}$. Simulation time step $\Delta t=0.01-1\mu s$.

\section{Results}
\subsection{Two characteristic time scales and their different dependence on the motor speed}
In Fig. 2(a), a typical case of time dependence of the rotor angle $\theta(t)$ from our model is shown. The motion of the rotor consists of two alternating phases: moving and waiting. The moving phase occurs when the net force on the motor is positive (in the direction of motion). The waiting phase is when the system reaches mechanical equilibrium (net force equals zero) and the motions are driven by thermal fluctuation.
The dynamics of the motor can thus be characterized by the two time scales $t_m$ and $t_w$. The waiting-time $t_{w}$ is the time the rotor spends fluctuating around a equilibrium position, i.e., the bottom of the total potential $V_t\equiv\sum_i^NV(\Delta\theta_i)$. Once in the waiting phase, the rotor can only start to have a net motion when a stator jumps to break the force balance and thus shift the equilibrium position forward. The subsequent net motion of the rotor to reach the new equilibrium position takes $t_{m}$, which is defined as the moving-time. The definitions of $t_w$ and $t_m$ are shown in Fig. 2(b).

The dynamics of the motor depend on the load, higher load leading to slower speed. We study how the two scales $t_m$ and $t_w$ vary with the load or equivalently the speed of the motor (speed is chosen because of its direct measurability in experiments). We find that the two time intervals have very different dependence on the motor speed as shown in Fig. 2(c). The waiting-time interval is determined by independent chemical transitions, i.e, by a Poisson process with rate $k$, so we have: $\langle t_w \rangle\propto \langle k^{-1}\rangle$. Since $k$ varies between two constants $k_{+}$ and $k_{-}$ (except for extreme high load where $k=0$), the averaged waiting-time has only a weak dependence on motor speed as shown in Fig. 2(c). On the other hand, the average moving-time can be estimated as:
$\langle t_{m}\rangle
\approx \delta_m/\omega_m$,
with $\delta_m$ the average angular movement, $\omega_m\equiv \tau_m/(\xi_R+\xi_L)$ the average speed, and $\tau_m\equiv\langle -V'_t\rangle_m $ the average net torque in the moving phase. Increasing the load $\xi_L$ leads to a decrease of the speed $\omega_m$ and an increase of the moving-time. In addition, at lower speed, it is more likely for stator to jump in the middle of a moving phase before the system reaches its force equilibrium. These premature stator jumps effectively increase $\delta_m$ and further increase $\langle t_m\rangle$. These two factors lead to a strong dependence of $\langle t_m\rangle$ on the load (or the speed) as shown in Fig. 2(c). Besides the difference in their average values, the distribution functions for $t_m$ and $t_w$ are also different (See Fig. S2 in SI for details).

\subsection{The two regimes of the torque-speed curve}
In Fig. 3, the torque-speed curves calculated from our model for 8 different stator numbers are shown. Our model results closely resemble the observed torque-speed curves.
There is a plateau regime with almost constant (10\% decrease) torque from zero up to a large speed ($\approx 100 Hz$ for $N=8$), followed by a steep declining regime of the torque, all the way to zero at a speed of roughly $300 Hz$.  By using the two time scales $t_m$, $t_w$, and noting that the net torque is zero during the waiting phase of the motor, the time-averaged torque $\tau$ and speed $\omega$ can be estimated:
 \begin{equation}
 \tau\approx \frac{\langle t_m\rangle }{\langle t_m\rangle +\langle t_w\rangle }\tau_m,\;\;\omega\approx \frac{\delta_m}{\langle t_m\rangle +\langle t_w\rangle }.
 \label{twtm}
\end{equation}
The two distinctive regimes in the torque-speed curve can be understood intuitively within our model by the different dependence of $\langle t_m\rangle $ and $\langle t_w\rangle $ on the speed
shown in the last section.

In the low-speed (high-load) regime defined by $\langle t_m\rangle  \gg \langle t_w\rangle$, we have $\tau\approx \tau_m$ and $\omega\approx \delta_m/\langle t_m\rangle$ from Eq.(\ref{twtm}). As discussed in the last section, for low speed a stator can jump prematurely during the moving phase before the system reaches the bottom of the potential well. As a result, each stator spends most of its time generating positive torque $\tau_0$. Therefore, in this high-load regime, while the speed changes significantly, the torque stays near its maximum value $\tau_{max}= N\tau_0$, which is proportional to the number of stators.

In the high-speed (low-load) regime defined by $\langle t_m \rangle \ll \langle t_w \rangle$, we have $\tau\approx \tau_m\langle t_m \rangle/\langle t_w\rangle $ and $\omega\approx \delta_m/\langle t_w\rangle $ from Eq.(\ref{twtm}). As shown in Fig. 2(c), for increasing speed $\langle t_m \rangle$ decreases quickly while $\langle t_w\rangle$ remains roughly the same. This naturally explains the steep decrease of the torque $\tau$ with speed in the high-speed regime. Intuitively, in this high-speed regime, a stator can be pushed into the negative torque region ($\Delta\theta>0$) because the rotor rotates too fast for the premature jump to occur. As the stators spend large fractions of their time dragging the rotor, the torque of the motor decreases quickly.

The different dependence of the waiting and moving-time intervals on the speed not only gives a clear general explanation for the two regimes of the torque-speed curve, it also explains the sharpness of the transition between the two regimes. Since the dependence of $\langle t_{m}\rangle $ on the speed is much steeper than that of $\langle t_{w}\rangle $ (as shown in Fig. 2(c)), the crossover between the two regimes takes place in a small region of the speed values, thus making the two regimes in the torque-speed curve well defined, as found in both experiments and simulations of our model.

\subsection{Independence of the motor speed on the number of stators at near zero load}
At near zero load, our model shows that the motor moves with a roughly constant speed that is independent of the number of stators, as demonstrated in Fig. 3. Recent resurrection experiments using gold nano-particle (extremely low load) indeed showed such independence\cite{YB08}. The mechanism for this surprising behavior can be understood with our model. In the low-load regime, the motor spends most its time in the waiting phase where the net torque is zero. In our model with symmetric potential $V$, this force equilibrium is achieved by having on average half of the stators pulling the rotor and the other half dragging it. If we number the stators in Fig. 1(d) from left to right, the rotor's equilibrium position sits between the $N/2$'th and the $(N/2+1)$'th stators. This equilibrium state breaks under two possible scenarios: 1) One of the $N/2$ dragging stators jumps to the pulling side. This occurs with a probability rate $Nk_- /2$. 2) The $(N/2+1)$'th stator jumps and shifts the equilibrium to a position between the $N/2$'th and the $(N/2+2)$'th stator. This occurs with probability rate $k_+$.
The average distance between the new and the old equilibrium positions are $\delta_m(\approx \delta_0/N)$ and $\delta_m/2$ for scenario 1) and 2) respectively. The fundamental reason for the decrease in step size with $N$ is due to the high duty ratio as first recognized in \cite{SB95,SB96}. Similar step size reduction with $N$ was recently observed in kinesin-1 motor\cite{LRHD07}. The maximum speed $\omega_{max}$ near zero load is then: \begin{equation}\omega_{max}(N) \approx \delta_m \frac{Nk_-}{2}+\frac{\delta_m}{2} k_+\approx \frac{k_-\delta_0}{2}[1+(k_+/k_-)N^{-1}],\label{s_max}\end{equation} which only depends weakly on $N$, if $k_+/k_-\ll 1$.
The estimated maximum speed $\omega_{max}\propto k_-\delta_0/2$ makes sense as $\omega_{max}$ should be limited by the step size and the maximum stepping frequency of an individual stator.

We have studied the dependence of $\omega_{max}$ on the ratio $r\equiv k_+/k_-$ and $N$ by numerical simulations of our model. In Fig. 4(a), we show the torque-speed curves for $N=1$ and $N=8$ for two different values of $r$: $r=0.2$ and $r=1.2$. To quantify the dependence of $\omega_{max}$ on $N$, we define a quantity $\Delta\equiv 2(\omega_{max}(1)-\omega_{max}(8))/(\omega_{max}(1)+\omega_{max}(8))$ to characterize the relative difference between the maximum speeds for motors with one and eight stators. As shown in Fig. 4(b), $\omega_{max}$ is roughly independent of $N$, i.e., $|\Delta|<0.1$ as long as $r\le 0.5$. However, $\omega_{max}(1)$ becomes significantly bigger than $\omega_{max}(8)$ for $r\ge 1$. The observed dependence of $\Delta$ on $r$ agrees well with the analytical estimate given by Eq.(\ref{s_max}).

\subsection{Motor speed fluctuation at different load levels and the estimate of step numbers}
The measured motor speed fluctuates due to two main factors: the external noise such as the Brownian noise and measurement noise, and the intrinsic probabilistic stepping dynamics of the stators. Samuel and Berg\cite{SB96} first investigated the speed fluctuations by studying the smoothness of the periodic motor motion characterized by
$\Gamma\equiv n \langle T_{1}\rangle ^{2}/(\langle T_{n}^2\rangle-\langle T_{n}\rangle ^{2})$,
where $T_{n}$ is the period for $n$ revolutions.
By measuring $\Gamma$ in a resurrection experiment where the stator number is inferred from the discrete increments in average motor speed, it was found that $\Gamma$ is proportional to the number of stators. The proportionality constant was interpreted as the number of steps per revolution.
Here, we analyze the motor fluctuation by using our model to understand how different noise sources contribute to $\Gamma$ and how $\Gamma$ behaves differently at different load levels.

For low loads, the motor spend most of its time in the waiting phase. The average motor step size is $\delta_m\approx \delta_0/N(\ll 2\pi)$, there are $n_s\equiv 2\pi/\delta_m\approx 2\pi N/\delta_0$ steps in each revolution, and the average periodicity is $\langle T_1\rangle =n_s\langle t_w\rangle $. Since the waiting-time intervals are uncorrelated, the variance of the $n-$revolution periodicity can be expressed as: $\langle T_n^2 \rangle -\langle T_n\rangle ^2=nn_s(\langle t_w^2\rangle -\langle t_w\rangle ^2)$. Furthermore, because the waiting-time $t_{w}$ is determined by a Poisson process, its variance is equal to $\langle t_{w}\rangle ^{2}$. $\Gamma$ can thus be written as:
\begin{eqnarray}\label{bergnumber2}
\Gamma\equiv \frac{n\langle T_1\rangle ^2}{\langle T_n^2\rangle -\langle T_n\rangle ^2}\approx \frac{n_s \langle t_{w}\rangle ^{2}}{\langle t_{w}^2\rangle-\langle t_{w}\rangle ^{2}}\approx \frac{2\pi}{\delta_0}N,
\end{eqnarray}
showing that $\Gamma=\gamma N$ is proportional to the stator number $N$, and the proportionality constant $\gamma=2\pi/\delta_0$ corresponds to the number of steps per stator per revolution. This behavior is verified in our model by calculating $\Gamma$ during a simulated resurrection process, where additional stators are added by a Poisson process with time constant $t_s=400s$ (Fig. 5). For near zero load, the average speed is independent of the stator number in agreement with \cite{YB08} (see Fig. 5(a)). However, $\Gamma$ increases by a fixed amount $\gamma=2\pi/\delta_0$ as a new stator is incorporated into the system as shown in Fig. 5(b), consistent with the analytical result by Eq.(\ref{bergnumber2}). The behavior of $\Gamma$ as shown in Fig. 5(b) represents a quantitative prediction of our model that could be tested in resurrection experiments with extreme low load, such as in \cite{YB08}.

For high load, the net torque is roughly constant $\tau\approx  N\tau_0$ and the speed can be expressed as $\omega_0=\tau/(\xi_L+\xi_R)\approx N\tau_0/\xi_L$, which explains the constant increment of speed for every additional stator (up to eight) seen in our model (Fig. 5(c)) as well as in the resurrection experiments by Blair and Berg\cite{BB88}. For additional stators beyond a certain large number of stators, the speed passes the knee in the speed-torque curve and our model predicts a decrease in the speed increment, which is consistent with the recent experiments by Reid et al \cite{RLCLAB06} that showed the same decrease in speed increment as stator number goes up to $N=11$. The dynamics of the load angle can be obtained by summing Eqs.(1-2) and taking the limit $\xi_R/\xi_L\rightarrow 0$. This leads to:
$\dot{\theta_L}=\omega_0+\sqrt{2K_BT/\xi_L}\beta(t)$,
which describes the simple motion of the load with a constant speed $\omega_0$ perturbed by random noise. From the equation for $\theta_L$, the periodicity and its variance can be determined: $\langle T_n\rangle \approx 2n \pi/\omega_0$, $\langle T_n^2\rangle -\langle T_n\rangle ^2\approx 4n\pi k_BT/(\xi_L\omega_0^3) $.  We can now express $\Gamma$ as:
\begin{equation} \Gamma\equiv \frac{n\langle T_1\rangle ^2}{\langle T_n^2\rangle -\langle T_n\rangle ^2}\approx \frac{\pi \xi_L}{k_BT}\omega_0\approx \frac{\pi\tau_0}{k_BT}N.\end{equation}
$\Gamma$ is again proportional to $N$ through its dependence on the speed $\omega_0$. However, unlike in the low-load regime, the proportionality constant $\gamma=\pi\tau_0/(k_BT)$ has nothing to do with the number of steps per revolution. Instead, $\gamma$ depends on the ratio between the intrinsic driving force ($\tau_0$) and the external noise $k_BT$, as the driving force overcomes the external noise to make the motor moves smoothly. This behavior is verified in our model by calculating $\Gamma$ during the resurrection simulation. As shown in Fig. 5(d), $\Gamma$ goes up with the number of stators but with a much larger proportionality constant $\gamma$, which quantitatively agrees with the expression $\pi\tau_0/(k_BT)$ from our analysis.

Therefore, although the motor-speed fluctuation is always suppressed by higher numbers of stators, the mechanisms are different for different load levels. For low load, the smoother motion for larger $N$ is caused by the increase in step number per revolution. For high load, the smoother motion for larger $N$ is caused by larger driving force (therefore larger speed) in comparison with the constant external noise. The difference in motor fluctuation between the high and the low load regimes is confirmed by our simulation as shown in Fig. 6, where the proportional constant $\Gamma$ is shown for different values of external noise strength $k_BT$ (Fig. 6(a)) and different load (Fig. 6(b)).

\section{Summary and Discussion}
We have presented a mathematical description of the rotary flagellar motor driven by hand-over-hand power-thrusts of multiple stators attached to the motor.  All key observed flagellar motor properties\cite{B03}, including those from a recent resurrection experiment at near zero load\cite{YB08}, can be explained consistently within our model. The crucial ingredient of our model is that the hand-switching rate depends on the force between rotor and stator. This feature is known to be valid for other molecular motors, including kinesin\cite{SMBlock99} and myosin\cite{OMOOCI07}. Therefore our model should be generally applicable to the study of these linear motors, especially in the case when there are multiple power-generating units attached to the same track\cite{LRHD07}.

For the flagellar motor, we find that its dynamics follows an alternating moving and waiting pattern characterized by two time scales $ t_m$ and $t_w $. The mechanism underlying 
the observed torque-speed relationship and its dependence on the number of stators, is revealed by studying the dependence of these two time scales on the load. 
For high load, $\langle t_w\rangle \ll \langle t_m\rangle $, the motor spend most time moving (albeit slowly) with all the stators pulling the motor in the same direction. So
the torques generated by individual stators are additive, leading to a roughly constant torque $\tau_{max}\approx N\tau_0$, which persists up to the knee speed $\omega_n$. Microscopically, the existence of this torque-plateau regime is due to the premature stator jumps which prevent the stators from going into the negative torque region.
Since the rate of the premature jumps is $k(\Delta\theta <0)$, larger $k_{+}$ and larger cutoff $\delta_c$ increase the knee-speed $\omega_n$ (see Fig. S3 in SI for details). For low load, $\langle t_w\rangle \gg\langle t_m \rangle $, the motor spends most time in the waiting state.
A waiting period ends when one of the dragging stators jumps to the pulling side or the pulling stator closest to the bottom of the potential well jumps. Therefore the maximum motor speed $\omega_{max}$ is limited by the maximum jumping rate of the stators. $\omega_{max}$ can be estimated from our model.  Eq.(\ref{s_max})
shows that $\omega_{max}$ has only a weak dependence on $N$ for small $k_{+}/k_{-}$, as confirmed by simulations of our model, and in agreement
with recent resurrection experiments at near zero load\cite{YB08}. Eq.(\ref{s_max}) also explains the strong dependence of $\omega_{max}$ on $N$ in a recent model by Xing et al\cite{XBBO06}. The jumping probability used in \cite{XBBO06} has a complicated profile and is maximum in the positive torque region. In our model, this would correspond to having $k_{+}/k_-\gg 1$, which is the opposite to what is required to achieve independence of $\omega_{max}$ on $N$.

The robustness of our results were verified using different forms of the rotor-stator potential $V$ and the force function $F$ between the load and the rotor. In particular, we have studied a smoothed symmetric potential with a parabolic bottom and an asymmetric potential $V$ (similar to the one used in \cite{XBBO06}) where the negative torque ($\tau_{-}$) is bigger than the positive torque ($\tau_{+}$). We find that all of our general results remain the same (see SI and Fig. S4\&S5 for details). For the asymmetric potential, the condition for $\omega_{max}$ being independent of $N$ is generalized to $k_{+}/k_{-}\ll \tau_{+}/\tau_{-}$. From the analysis and direct simulation of our model, we do not find any significant dependence of the torque-speed curve characteristics on the specifics of the force function $F$ between the load and the rotor. In particular, contrary to what was proposed in \cite{XBBO06}, there is no difference between the case of a viscous load that interact with the rotor through a soft spring and a viscous load without spring (see Fig. S6 in SI for details).

Besides the torque-speed curve which describes the time averaged behavior of the motor, we have also studied the speed fluctuation for individual flagellar motor. We find that the fluctuation is damped by the number of stators for all load levels. However, we show that the dominating source of the motor fluctuation is different depending on the load. For low load, the speed fluctuation is dominated by the discrete stochastic stepping events whereas for the high load, it is controlled by the external noise, such as Brownian fluctuations or possibly measurement noises. The original measurements on motor fluctuation by Samuel and Berg\cite{SB96} were done in the high-load regime as evidenced by the discrete increment of speed in their resurrection experiment. Therefore the strength of the fluctuations obtained there is probably more reflective of the strength of external noise than the number of steps per revolution. It would be interesting to perform the fluctuation analysis in the low-load regime as achieved in \cite{YB08} to determine the steps number per revolution and compare with the recent direct observation of the steps\cite{SRLYHIB05}.

Simple relations between the macroscopic observables ($\tau_{max}$, $\omega_n$,  $\omega_{max}$) and the microscopic variables of the system ($\tau_0$, $k_+$, $k_-$) are established by analysis of our model. These relations can be used to predict the microscopic parameters quantitatively from the torque-speed measurements. They can also be used to study the dependence of the flagellar motor properties on other relevant external parameters such as the \emph{pmf}, the temperature, and solvent isotope effects. For example, since changing of \emph{pmf} gives rise to self-similar torque-speed curves\cite{GB03}, we conclude from our model that larger \emph{pmf} not only increases the chemical transition rates $k$'s, it also increases the stator-rotor interaction strength $\tau_0$. Changing temperature or replacing $H^{+}$ with $D^{+}$ (solvent isotope effect) should affect the chemical transition rates. These changes in $k_{+}$ and $k_{-}$ lead to changes in the knee speed $\omega_n$ and the maximum speed $\omega_{max}$ in our model  without changing the maximum torque at stall, which is consistent with previous experimental observations\cite{BT93,CB00b}.

Backward stator jumps with $\theta^S\rightarrow \theta^S-\delta_0$ can be incorporated in our model to study the relatively rare motor back-steps\cite{SRLYHIB05}. The back-jumps are neglected in this paper as their probabilities are much smaller than those for the forward jump in the region of relative angles ($\Delta\theta >-\delta_c$) relevant for our study here. However, we expect the back-jumps to become dominant for $\Delta\theta <-\delta_c$ where the forward jumps are prohibited. Since the landing points of these back-jumps are still on the positive side of the potential with positive torque $\tau_0$, inclusion of back-jumps in our model for $\Delta\theta<-\delta_c$ can naturally explain the observed torque continuity near stall when the motor is driven backwards by an optical tweezer\cite{BB97}.

In our model, the step size depends inversely on the stator number $N$. This behavior is a general consequence of duty ratio being unity and independent stepping of the stators, as pointed out by Samuel and Berg\cite{SB96}. This $N^{-1}$ dependence of the step size seems to be inconsistent with an ``apparent independence" of step size on $N$ claimed in \cite{SRLYHIB05}. However, a careful study of the experimental data reveals that the $N$ dependence of the step size can not be ruled out, because the step size distribution was measured for a varying population of stators, whose number was neither controlled nor measured precisely in \cite{SRLYHIB05}. An unambiguous way to determine whether the step size depends on $N$ is to measure the step size for different fixed $N$ or at least to measure $N$ simultaneously. Such experiment was done recently for kinesin-1\cite{LRHD07} and showed that step size for $N=2$ is half of that for $N=1$.

Our model works for the clockwise (CW) as well as for the counterclockwise (CCW) rotation. It was recently suggested that the switching between the CW and CCW state of the motor is a non-equilibrium process
and the energy needed to drive the motor switch could be provided by the same \emph{pmf} that drives the mechanical motion of the motor \cite{T08}. The possible link between the switching process and the rotational motion of the motor is supported by experimental observations \cite{FRB03} showing that the average switching frequency depends on the proton flux. It is therefore highly desirable to develop an integrated model to describe both the mechanical part of the flagellar motor, associated with the rotational motion, with the signaling part, associated with the switching process. More experimental information on the components of the motor (M-ring/C-ring/MotAB) and how they interact with each other are needed to achieve this goal.

\section{acknowledgments}
This work is partially supported by a NSF grant (CCF-0635134) to YT.

\end{singlespace}
\begin{figure*}[!t]
\vspace{2cm}
\centerline{\includegraphics[angle=0,width=0.6\textwidth]{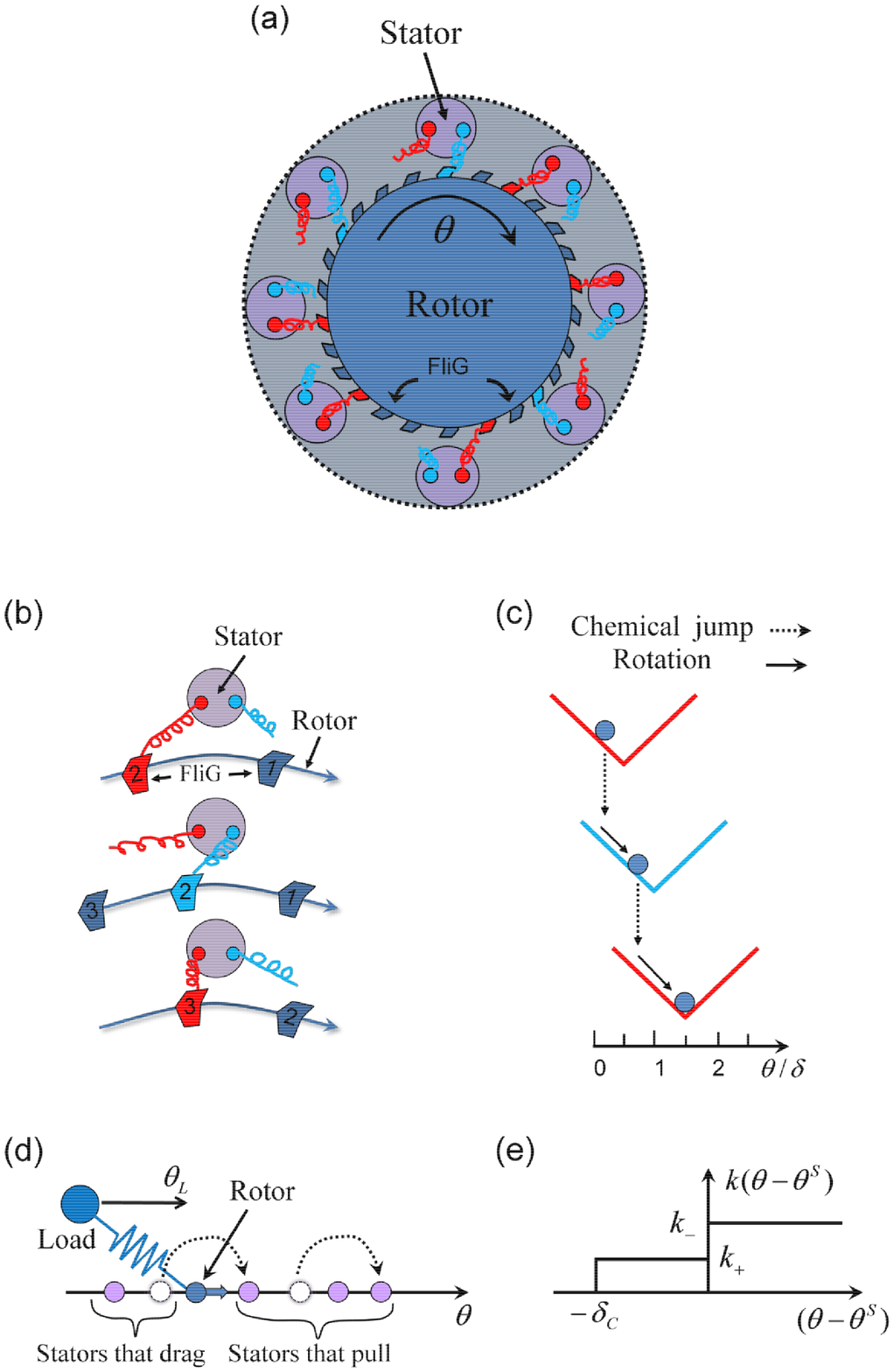}}
\caption{A Model for the flagellar rotary motor. (a) Schematic illustration of the rotor-stators spatial arrangement. The rotor contains 26 FliG proteins and there are multiple stators, each with two subunits (red and light blue springs). (b) A sequence of three rotor-stator configurations (from top to bottom) illustrating the hand-over-hand interaction between the two subunits of a stator and the FliG proteins in the rotor. (c) The same sequence as in (b) is shown in the potential landscape. The solid arrow represents the physical rotation of the rotor angle ($\theta$) down a given (V-shaped) potential, the dotted arrow represents the chemical change (switching of hands) that shifts the potential. (d) The full motor model with multiple stators in the angle space. Each stator is represented by its internal angle $\theta^S$. The rotor is pulled forward by the  stators in front it and dragged back by the stators behind it. The stator angle can only change by jumping forward with rate $k$ that depends on the relative angle $\Delta\theta=\theta-\theta^{S}$. The form of $k(\Delta\theta)$ used in this paper is given in (e), which shows the dragging stators have a higher jump rate $k_->k_+$ and a cutoff angle $-\delta_c$ where $k(\Delta\theta<-\delta_c)=0$.}
\end{figure*}
\vspace{2cm}
\begin{figure*}
\vspace{2cm}
\centerline{\includegraphics[angle=0,width=0.6\textwidth]{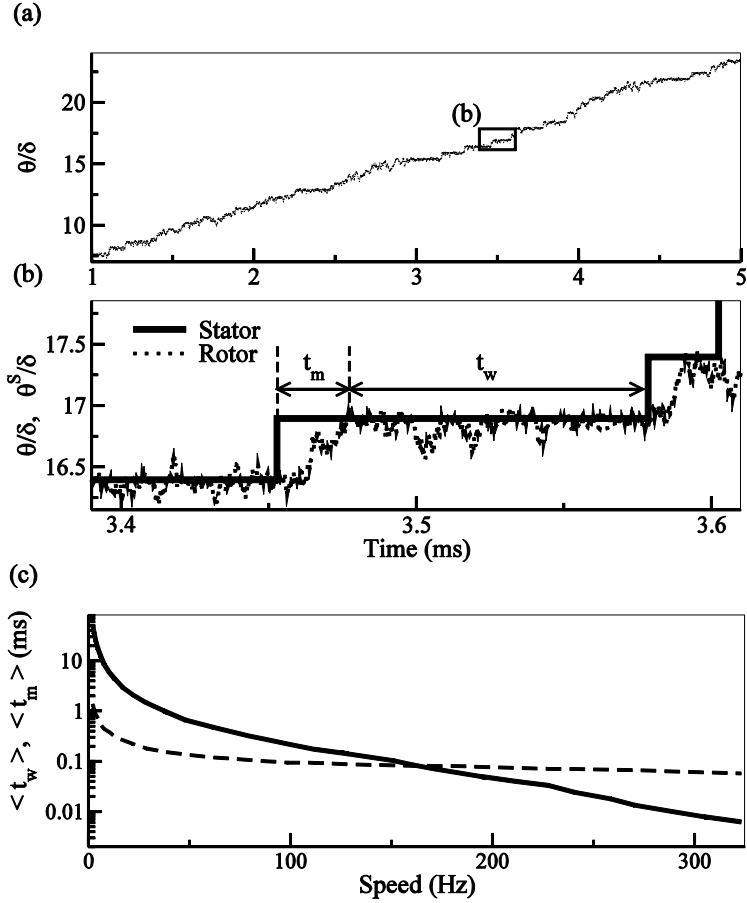}}
\caption{The motor dynamics and its dependence on the load. (a) Rotor angle $\theta$ versus time for $\sim 4/5$ of a revolution for $N=1$. The angular unit is the FliG periodicity $\delta$. The insert is enlarged in (b). (b) Zoom of the time series in (a) showing two complete steps. Solid line shows the stator position $\theta^{S}$. A jump in $\theta^S$ marks the start of a moving phase for the rotor and the waiting phase starts when the rotor catches up with the stator. The definitions of the moving-time $t_m$ and the waiting-time $t_w$ are shown. (c) The average waiting-time $\langle t_w\rangle $ (dashed line) and the average moving-time $\langle t_m\rangle $ (solid line) over 500 revolution as a function of the rotational speed for N$=1$.  $\langle t_w\rangle $ decreases slowly with increasing speed from 1ms to 0.1ms while $\langle t_m\rangle $ decreases much faster from roughly 50ms to 0.005ms.}
\end{figure*}
\vspace{2cm}
\begin{figure*}
\vspace{2cm}
\centerline{\includegraphics[width=0.6\textwidth]{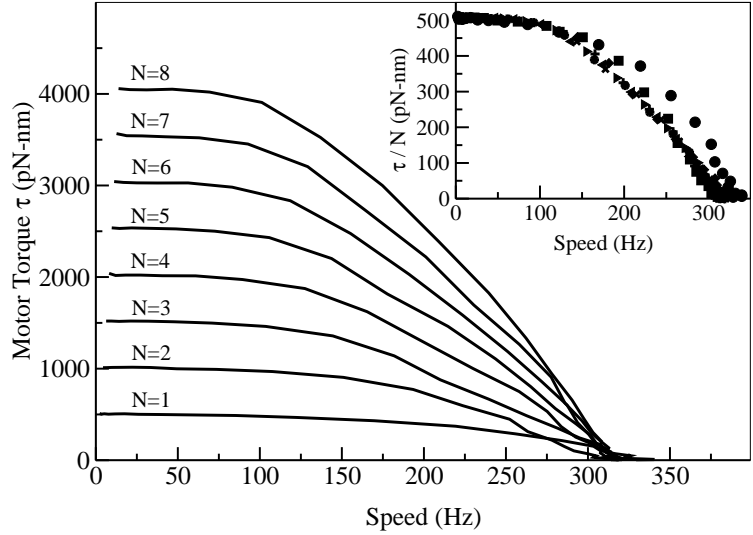}}
\caption{The torque-speed ($\tau-\omega$) curves for different stator numbers ($N=1$ to $N=8$) from our model. Two regimes of the $\tau-\omega$ curves, i.e., constant $\tau$ up to a large knee speed $\omega_n$ and fast decrease of $\tau$ to zero at the maximum speed $\omega_{max}$, are evident for all stator numbers. The torque per stator $\tau/N$ versus the speed $\omega$ is shown in the insert. The torque at stall scales with $N$ while the maximum speed is independent of $N$.}
\end{figure*}
\vspace{2cm}
\begin{figure*}
\vspace{2cm}
\centerline{\includegraphics[width=0.6\textwidth]{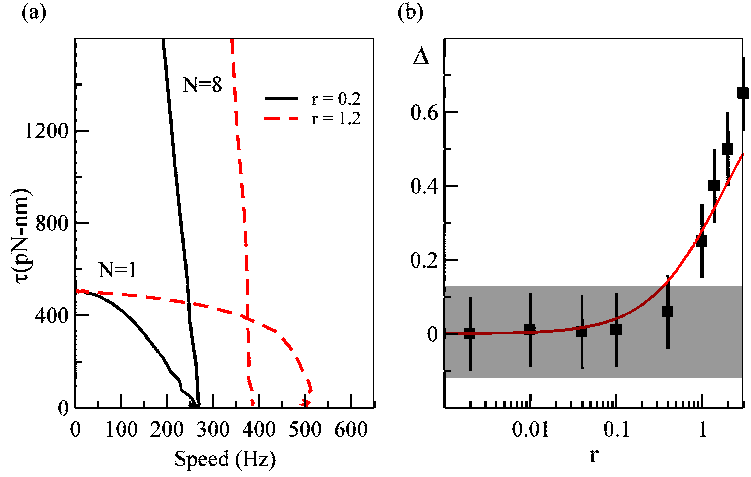}}
\caption{Dependence of the maximum speed at zero load on $N$ as a function of $r\equiv k_{+}/k_{-}$. (a) Torque-Speed curves for $r=0.2$ (solid lines) and $r=1.2$ (dotted lines) are shown for $N=1$ (black lines) and $N=8$ (red lines). We change $r$ by varying $k_{+}$ and keeping $k_{-}$ constant. For $r=0.2$, the maximum velocities $\omega_{max}(1)$ and $\omega_{max}(8)$ at zero load are roughly the same, while they differ significantly for $r=1.2$. (b) The dependence of   $\Delta=2(\omega_{max}(1)-\omega_{max}(8))/(\omega_{max}(1)+\omega_{max}(8))$ on r. The red line represents our analytical predictions from Eq.(5). The shaded region shows the $\sim 12\%$ experimental error.}
\end{figure*}
\vspace{2cm}
\begin{figure*}
\vspace{2cm}
\centerline{\includegraphics[width=0.6\textwidth]{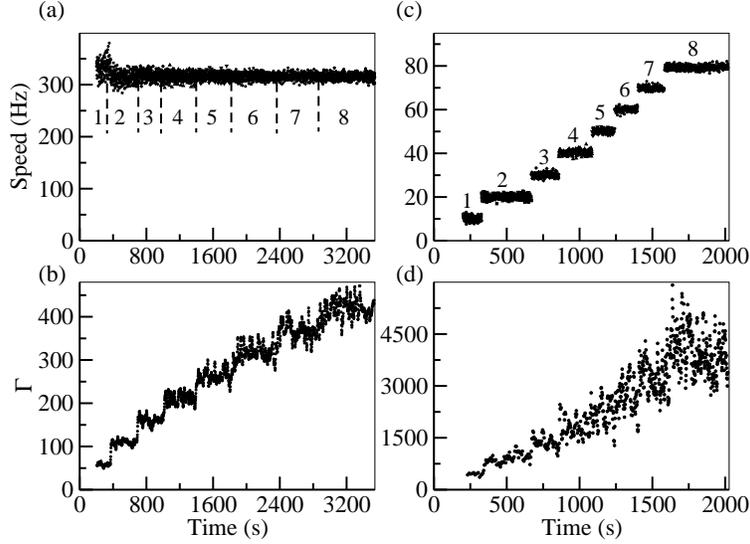}}
\caption{The motor speed and its fluctuation in a simulation of the resurrection process for low and high loads. (a) Speeds as a function of time after successive stators are added at low load $(\xi_L=0.002$pN-nm-s-rad$^{-1})$. The simulation shows no dependence of the speed on the number of stator (labeled with a number form $1$ to $8$). (b) The smoothness parameter $\Gamma$ with n$=5$ calculated from the time series shown in (a). $\Gamma$ value is calculated with a moving-time window of $1.5s$ ($\sim 500 $ revolutions). (c) Same as (a) at high load $(\xi_L=8$pN-nm-s-rad$^{-1})$showing the roughly linear dependence of the motor speed $\omega$ on $N$. (d) $\Gamma$ at high load from the time series shown in (c) using a window of roughly 10$s$ corresponding to a total of 100 revolutions. (b) and (d) show that $\Gamma$ increases with $N$ at both high and low load.}
\end{figure*}
\vspace{2cm}
\begin{figure*}
\vspace{2cm}
\centerline{\includegraphics[width=0.6\textwidth]{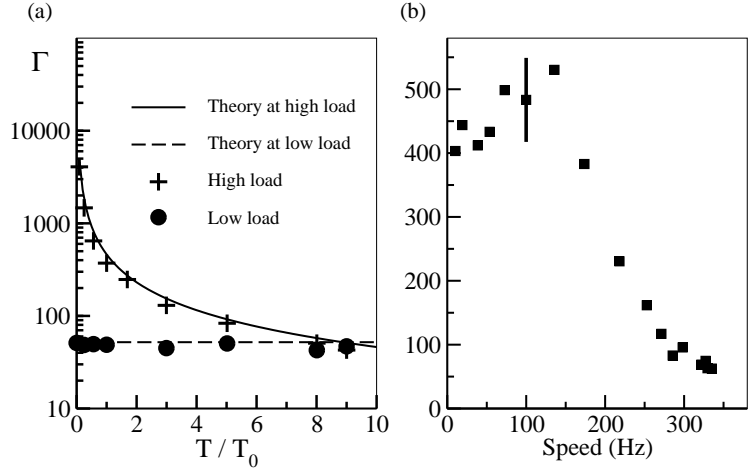}}
\caption{The speed fluctuation and its dependence on the load (with $N=1$). (a) The dependence of $\Gamma$ on the external noise strength defined as $T/T_0$, where $T_0$ is the room temperature. $\Gamma$ depends strongly on $T$ for high load (crosses) while it is a constant determined by the step size at load load (dots), in agreement with our theoretical results (lines).(b) The dependence of $\Gamma$ on speed $\omega$ for $T=T_0$. Typical error bar (SD) is shown at high load; the error at low load is comparable to the symbol size.}
\end{figure*}
\end{document}